\documentclass[a4paper,11pt]{article}

\renewcommand{\thetable}{\Roman{table}}

\pdfoutput=1 

\usepackage{jheppub} 
\usepackage[T1]{fontenc} 
\usepackage{lineno,hyperref}
\usepackage{caption}
\usepackage{subcaption}
\usepackage{soul}
\title{\boldmath A study of multicavity concept applied to hexagonal coaxial haloscopes}

\author[*,a,b]{J.M.~Garc\'ia-Barcel\'o,}
\author[c]{Jose R. Navarro-Madrid,}
\author[c]{and Alejandro\:Díaz-Morcillo}

\affiliation[a]{Center for Astroparticles and High Energy Physics (CAPA), Universidad de Zaragoza, 50009 Zaragoza, Spain}
\affiliation[b]{Max-Planck-Institut f\"{u}r Physik (Werner-Heisenberg-Institut), Boltzmannstraße 8, 85748 M\"{u}nchen, Germany}
\affiliation[c]{Department of Information Technologies and Communications, Universidad Politécnica de Cartagena, 30202 - Cartagena, Spain}

\affiliation[*]{Corresponding author}

\emailAdd{josemaria.garcia@unizar.es}

\abstract{In this work, a study on the development of scalable multicavity architectures for axion haloscopes, based on a hexagonal coaxial geometry operating at $30$~GHz frequencies, is presented. To enhance the scanning rate and sensitivity within the limited volume of experimental magnet bores, the transition from a baseline single-cavity design to dual and triple-subcavity configurations is investigated. A novel tuning mechanism based on the rotation of one or two inner hexagonal prisms is implemented, providing a robust method to shift the resonant frequency while maintaining high form and quality factors. The results show that the triple-subcavity design achieves an improvement of $\times3$ over the single-cavity baseline. The scaling potential of quad-subcavity architectures under a strict radial constraint of $25$~mm is further explored. Theoretical analysis confirms that a four-subcavity system is feasible within a certain radial clearance, provided that wall thicknesses are strategically optimised to ensure frequency stability. The practical challenges associated with mode splitting, manufacturing tolerances, and thermal management in these high-order systems are also discussed. This one-port multicavity approach offers a viable path toward increasing the sensitive volume of haloscopes, enabling more efficient exploration of the axion dark matter parameter space in high-frequency regimes.}

\begin{document}
\maketitle
\flushbottom

\section{Introduction}
\label{s:Introduction}

Axions are hypothetical pseudo-Goldstone particles that were proposed by Weinberg \cite{weinberg1978} and Wilzcek \cite{Wilczek1977} as a solution to the strong-CP problem in Quantum Chromodynamics (QCD) \cite{Peccei1977,Pecceib1977} and whose expected characteristics make this particle difficult to detect, since it would be extremely light, neutral and are predicted to interact only minimally with ordinary matter and electromagnetic (EM) fields. These characteristics make them also great candidates to be part of Dark Matter (DM). Recently, there has been a significant rise in the number of experiments and methods aimed at either identifying or ruling out axions as DM (see, for example, \cite{IGOR2018}).

An axion haloscope is a type of experimental device designed to detect DM axions, and whose aim is to detect these particles by converting them into observable EM signals (i.e., photons) through the inverse Primakoff effect in the presence of a strong magnetic field \cite{Primakoff1951}. To enhance this conversion method, axion haloscopes employ resonators with high quality factors, such as microwave cavities \cite{Sikivie1983}. The conversion from these particles to photons occurs when the axion mass closely aligns with the resonant frequency of the detector. Since the mass of the axion is unknown, and it is directly related to the generated photons frequency ($f \approx m_a c^2 / h$), different kinds of haloscopes have been designed in order to cover distinct frequency ranges.

Currently, notable examples of haloscope experiments are ADMX (Axion Dark Matter eXperiment) \cite{ADMXstatus} and CAPP (Center for Axion and Precision Physics) \cite{CAPP_2023}, which operate below and above $1$~GHz, respectively. The ADMX collaboration has reached significant sensitivities in the $2.81-3.31$~$\mu$eV range \cite{ADMXa2021}, while CAPP has proposed a DFSZ-sensitive haloscope exploring the $4.51$ to $4.59$~$\mu$eV range \cite{CAPP_2023}. The HAYSTAC team has achieved KSVZ-sensitive results in the $16.96-17.28$~$\mu$eV and $23.15-24.0$~$\mu$eV mass ranges \cite{HAYSTAC2018,HAYSTAC2021}. Other groups, such as RADES \cite{RADES_paper3,RADES_BabyIAXO} (which this study is part of), and QUAX \cite{QUAX_2021,QUAX_Q} have also made notable advances in haloscope technology.

The main objectives of an efficient axion detection system are to maximise the power produced by the axion-photon interaction and to broaden the spectrum of axion masses being probed (i.e., the range of scanned frequencies). The detected radiofrequency (RF) power ($P_d$) depends on the intrinsic properties of the axion and the experimental cavity detector, as described in \cite{RADESuniverse} (in natural units):
\begin{equation}
\label{eq:Pd}
    P_d = \kappa g_{a\gamma}^{2}\frac{\rho_a}{m_a}B_{e}^{2}CVQ_l,
\end{equation}
where $\kappa$ is the external receiver coupling (which is commonly re-expressed in terms of the coupling coefficient $\beta=\frac{\kappa}{1-\kappa}$), $g_{a\gamma}$ the axion-photon coupling, $\rho_a$ the DM density, $m_a$ the mass of the axion, $B_e$ the external static magnetic field, $C$ the form factor, $V$ the volume of the cavity, and $Q_l$ its loaded quality factor. To have critical coupling, $\kappa = 0.5$ is needed. In contrast to the loaded quality factor, the unloaded quality factor ($Q_0$) is a widely used parameter for characterising resonant cavities, as it does not depend on input/output couplings ($\kappa$) \cite{Pozar,Rezaee_2012}. The form factor $C$ quantifies the alignment between the external magnetostatic field of the employed magnet ($\vec{B}_e$) and the electric field generated by the axion field after the conversion to photons in the cavity ($\vec{E}$), and it is defined as
\begin{equation}
\label{eq:C}
    C \, = \, \frac{|\int _V \, \vec{E} \cdot \vec{B}_e \, dV|^2}{\int_V \, |\vec{B}_e|^2 \, dV \int_V \, \varepsilon_r \, |\vec{E}|^2 \, dV},
\end{equation}
where, $\varepsilon_r$ is the relative electrical permittivity in the cavity.

The magnetic field $B_e$ will depend on the magnet used for the axion search, which will also limit the experimental space and, thus, the maximum dimensions of the resonant cavity. The magnets most commonly used in haloscope experiments are of solenoid type, whose bore section is circular, making the cylindrical shape the most efficient geometry for designing the resonant cavities. However, there are cases where dipole magnets are used, so a rectangular cavity can also be efficient.

To increase the likelihood of detecting axions, the haloscope must be tuned to the resonant frequency corresponding to the expected axion mass range under investigation. Consequently, the scanning rate ($\frac{dm_a}{dt}$) is defined as the speed at which a mass range can be scanned at a given sensitivity $g_{a \gamma}$ \cite{CAPP_revisiting}, and is commonly used to extract the haloscope Figure of Merit (FoM) that takes into account the cavity design parameters:

\begin{equation}
\label{eq:FoM}
    \text{FoM} = Q_0 V^2 C^2.
\end{equation}

From equation~\ref{eq:C}, it can be extracted that the more parallel the electric field of the resonant mode excited (by the axion field) in the cavity is to the magnetic field of the magnet, the higher the form factor will be. For a solenoid magnet, this is the case of the $\mathrm{TM_{010}}$ mode, which is most commonly used in haloscope experiments. The diameter of the cylindrical cavity determines the resonant frequency of this mode, and consequently, the axion mass range to which the experiment is most sensitive, and probing higher axion masses (that is, higher frequencies) typically requires smaller diameters. However, reducing the cavity size leads to a decrease in the signal power and the scanning rate, as the former scales as $Q_0V$ (see eq.~\ref{eq:Pd}) and the latter as $Q_0V^2$ (see eq.~\ref{eq:FoM}).

As a result, in recent years, new techniques have been investigated to explore the high-mass region. One of the techniques consists of employing multicavity structures, as in RADES \cite{RADES_paper1,RADES_paper2} and CAPP \cite{CAPP_multicav,CAPP_Youn_2024}, which provides promising results for high-frequency haloscope volume enhancement. This technique enables having multicavities in more than one dimension, both in rectangular and cylindrical geometries (see \cite{VolumeRec} and \cite{VolumeCyl}, respectively). Also, the use of dielectric cavities \cite{MADMAX:2019pub}, multirod cavities \cite{10.1063/5.0016125}, wire metamaterial cavities (or plasma haloscopes) \cite{Plasma_haloscope}, or the multiple-cell structure \cite{Multiple_CAPP_study} are examples of other techniques that deal with the problem of frequency scaling. The main disadvantage of the multicavity technique lies in the difficulty of applying mechanical tuning systems in them due to the number of partitions and the precision required for the mechanical systems to perform accurate tuning in each subcavity. For example, the impact of the misalignment of a tuning system on a multicavity haloscope can be seen in \cite{10.3389/fphy.2024.1372846}.

Some recent approaches to tuning have emerged. The first one consists of thin shell cavities \cite{Shell_Kuo_2020,Shell_Kuo_2021,Shell_Kuo_2024} that achieve large volumes enabling a wide tuning range just by varying the separation between shells. The second technique is based on a coaxial resonator, which consists of two nested polygonal prisms, with the cavity residing in the space between them. The frequency tuning is achieved by rotating the inner prism in relation to the outer one, changing the distribution of the EM field in the cavity and thus varying the resonant frequency of the $\mathrm{TM_{010}}$ mode. In \cite{QUAX_poly}, a complete study of different prisms, their performance as haloscopes and the case of a hexagonal coaxial cavity can be consulted.

From this concept of independent nested polygonal cavities, in this work, the application of the multicavity concept to a hexagonal coaxial structure is studied, and an attempt is made to solve the problem of multicavity tuning by using the inner prism rotation system, extracting the FoM of a single cavity and two multicavity designs (double and triple subcavity configurations). The main motivation for replacing multiple cavities with a multicavity is obtaining the signal in a unique port, avoiding in this way the difficult coherent sum of signals from a number of ports. The article is organised as follows: In Section \ref{s:From_single_to_multicavity}, the design principles are established by transitioning from a single hexagonal coaxial cavity to a multicavity approach. Section \ref{s:Performance_analysis} provides a performance analysis of coupled subcavities, comparing double and triple configurations in terms of FoM and spectral coverage. The tuning dynamics and port coupling strategies for optimal signal extraction are explored in Section \ref{s:Tuning_dynamics_and_port_coupling}. Then, Section \ref{s:Scaling_potential} discusses the scaling potential to higher-order architectures and the practical challenges involved. Finally, a summary of the work and the main findings are presented in Section \ref{s:Conclusions}.

\section{From single to multicavity: design principles} \label{s:From_single_to_multicavity}

\subsection{Baseline: single hexagonal coaxial cavity}\label{ss:SingleHexCoaxCav}

\begin{figure}[htb]
\centering
\begin{subfigure}[b]{0.32\textwidth}
         \centering
         \includegraphics[width=1\textwidth]{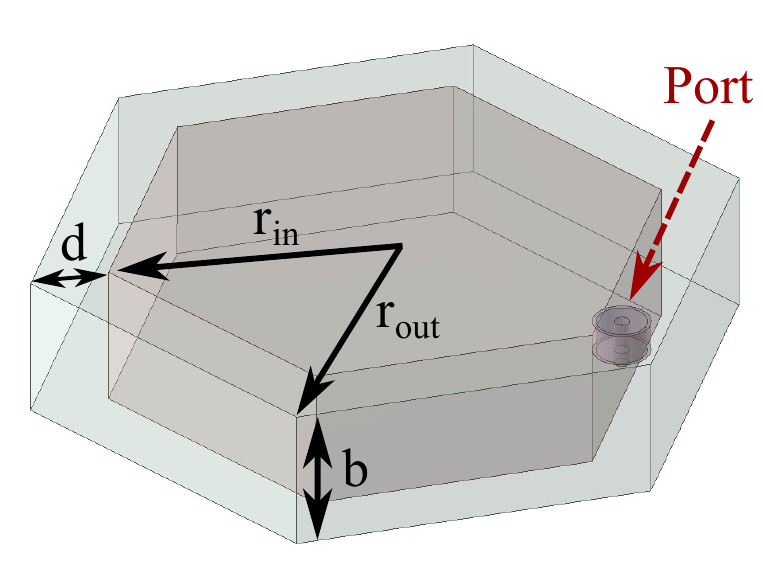}
         \caption{}
         \label{fig:3Dmodel_single}
\end{subfigure}
\hfill
\begin{subfigure}[b]{0.32\textwidth}
         \centering
         \includegraphics[width=1\textwidth]{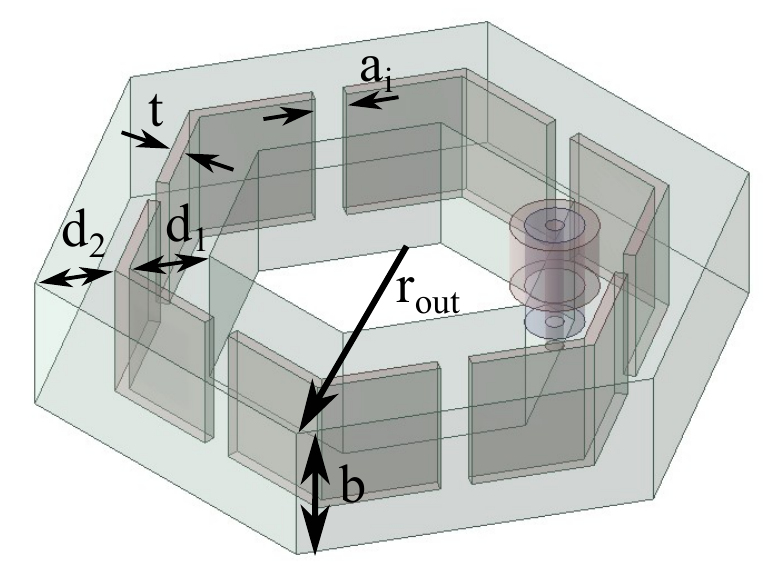}
         \caption{}
         \label{fig:3Dmodel_multicav_dual}
\end{subfigure}
\hfill
\begin{subfigure}[b]{0.32\textwidth}
         \centering
         \includegraphics[width=1\textwidth]{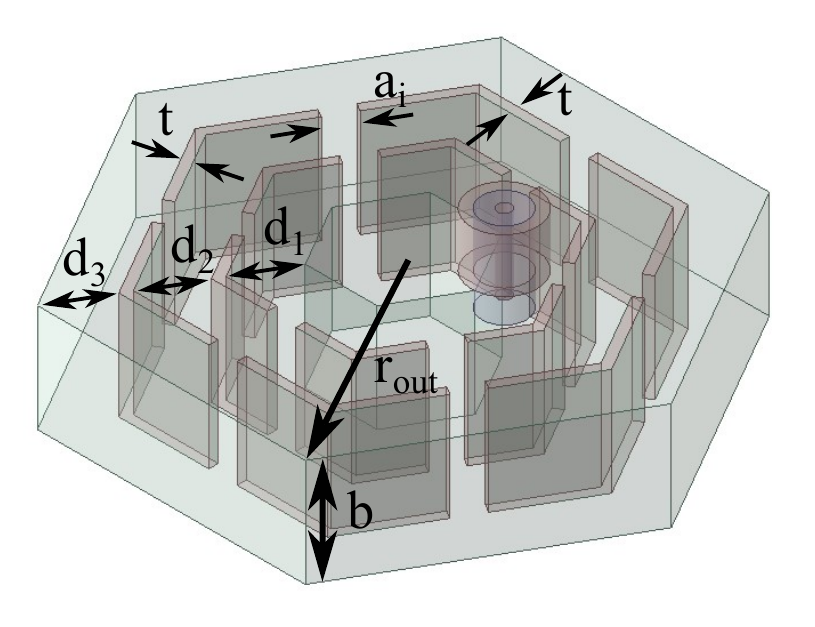}
         \caption{}
         \label{fig:3Dmodel_multicav_triple}
\end{subfigure}
\caption{3D models of different types of hexagonal resonant haloscopes: (a) single cavity, (b) double multicavity, (c) triple multicavity. A coaxial port is implemented at one subcavity for each haloscope.}
\label{fig:3Dmodel_single_and_multicavs}
\end{figure}

The geometry used in this study was originally proposed by the QUAX experiment in \cite{QUAX_poly}, which consists of two concentrically nested hexagonal prisms between the walls of which the cavity is located, as it is depicted in Figure~\ref{fig:3Dmodel_single}.
It is defined by two radii that correspond to the two prisms: $r_{in}$ for the inner one and $r_{out}$ for the outer one, being $r_{in} < r_{out}$. The gap between walls, that is, the cavity area, is defined by $d = r_{out}-r_{in}$, and gives approximately the resonant frequency of the $\mathrm{TM_{010}}$, since the geometry is not perfectly circular.

The tuning system proposed in \cite{QUAX_poly} is based on the rotation of the inner prism related to the outer one (defined by the angle $\phi$) to perturb the EM field distribution, leading to a break of the cylindrical symmetry and confining this field near the vertices. Related to this, the computation of the $C$ factor is compromised when using CST Eigensolver \cite{CST} due to the change of coupling between electric field lobes generated by rotating the inner prism when tuning, giving higher values of $C$ than the real ones. For that reason, the CST Frequency Domain solver has to be used in these simulations. Moreover, in \cite{QUAX_poly}, the effect of adding more sides to the prisms ($N$) at the frequency of the pseudo-$\mathrm{TM_{010}}$ mode for the initial position of the walls ($\phi = 0^\circ$), which is also related to the confinement of the electric field at the vertices, is described. The increase in the number of sides leads to a decrease in the cavity tuning range, as the coaxial cavity approaches a cylindrical shape. After an analytical study of possible polygonal cavity geometries, a compromise solution of using a hexagonal cavity was reached in \cite{QUAX_poly}. This cavity has been simulated, analysed, manufactured and measured, giving good results in terms of the FoM compared with other resonant cavity systems at around $10$~GHz.

In the case of the single coaxial cavity of this article, it has been started from the hexagonal design obtained in \cite{QUAX_poly}, scaling the dimensions for operating the pseudo-$\mathrm{TM_{010}}$ mode at $f_r = 30$~GHz, a frequency selected taking into account the experimental space of the solenoid magnet under analysis. The bore diameter of this magnet is $50$~mm, which restricts the maximum radius for the coaxial cavity to $r_{out} = 25$~mm, a value that has been adopted in this study for all designs, both single cavity and multicavity systems. The length of the haloscope designs explored in this work, $l$, is set as $l = 10$~mm to reduce the computational cost of the simulations. The signal is extracted by a coaxial probe located in one of the vertices at $d/2$, far enough away so that, when rotating the internal prism during tuning, it does not touch the walls. The model shown in Figure~\ref{fig:3Dmodel_single} has been designed by simulation in CST Studio Suite \cite{CST} with the Frequency Domain solver, giving an inner radius of $r_{in} = 19.3$~mm, and therefore, a cavity gap of $d = 5.7$~mm. With these dimensions, the cavity volume is $V = 6.658$~mL. Moreover, for all the simulations carried out in this article, the electrical conductivity for copper at $T=4$~K ($\sigma = 10^9$~S/m) is assumed, which results in a quality factor of $Q_0 = 23353$ for this single cavity. The obtained form factor is $C = 0.706$ and the resulting Figure of Merit is $\mathrm{FoM} = 0.516$~$\text{L}^2$.

\subsection{The multicavity approach: concept and geometry}\label{ss:Multicavity_approach}

There are several options to apply the multicavity concept to this design, depending on the type of coupling between subcavities, as explained in \cite{VolumeCyl}. In this case, in order to take advantage of the inner prism tuning system, an inductive-type radial coupling system has been chosen; that is, the irises for the subcavity couplings are located in the $\rho$ (radial) direction. An interresonator coupling $|k| = 0.025$ has been chosen, which avoids the mode clustering issue between configuration modes in multicavities \cite{VolumeCyl}. To maintain symmetry and achieve higher and better interresonator coupling, a configuration comprising an all-inductive iris system with one iris per side, located in the flat parts of the inner prisms, has been adopted. Some simulation tests were carried out using only one or two irises, and it was found that the coupling coefficient achieved was insufficient. The thickness of the inner walls separating subcavities has been set to $t = 1$~mm. In Figures~\ref{fig:3Dmodel_multicav_dual} and \ref{fig:3Dmodel_multicav_triple}, the polygonal multicavity haloscopes analysed in this work are depicted, based on 2 and 3 subcavities, respectively.

Unlike the single cavity design, in the case of multicavity systems, the tuning mechanism is based on the rotation of the walls separating the subcavities (one or two walls, depending on whether it is a double or triple configuration, respectively).

\section{Performance analysis of coupled subcavities} \label{s:Performance_analysis}

\subsection{Multicavity design results}\label{ss:Multicavity_design_results}

An optimisation process has been performed to set the pseudo-$\mathrm{TM_{010}^{++}}$\footnote{Here the superindices follow the notation in \cite{VolumeRec,VolumeCyl}, where '++' or '+++' indicates that the electric field gets a positive maximum in each subcavity. These are the mode configurations which yield the maximum form factor.} (double multicavity) and pseudo-$\mathrm{TM_{010}^{+++}}$ (triple multicavity) configuration modes to operate at $f_r = 30$~GHz (before tuning), giving the following resulting gaps (the cavity area): $d_1 = 5.6$~mm and $d_2 = 5.62$~mm for the inner and outer subcavities of the double multicavity, respectively, and $d_1 = 5.41$~mm, $d_2 = 5.32$~mm, $d_3 = 5.57$~mm for the inner, central and outer subcavities of the triple multicavity, respectively. Regarding the inductive irises, all of them were set with an iris width of $a_i = 2.2$~mm for the double configuration, and $a_i = 2.65$~mm for the triple configuration, to obtain the desired $|k|$ value mentioned in the previous section.

The achieved total volumes are $V_{2} = 11.127$~mL and $V_{3} = 13.705$~mL for the double and triple multicavities, respectively. This represents an increase of $\sim 70$~$\%$ and $\sim 106$~$\%$ for the double and triple multicavities, respectively, in relation to the single cavity design. At the point of $\phi=0^\circ$, the quality and form factors, as well as the Figure of Merit, were $Q_{0,2} = 20261$, $C_{2} = 0.673$ and $\mathrm{FoM}_{2} = 1.136$~$\mathrm{L}^2$, and $Q_{0,3} = 18765$, $C_{3} = 0.679$ and $\mathrm{FoM}_{3} = 1.625$~$\mathrm{L^2}$, for the double and triple multicavity configurations, respectively.

\subsection{Comparative analysis of FoM with frequency tuning}\label{ss:ComparativeFoM}

\begin{figure}[htb]
\centering
\begin{subfigure}[b]{0.4\textwidth}
         \centering
         \includegraphics[width=1\textwidth]{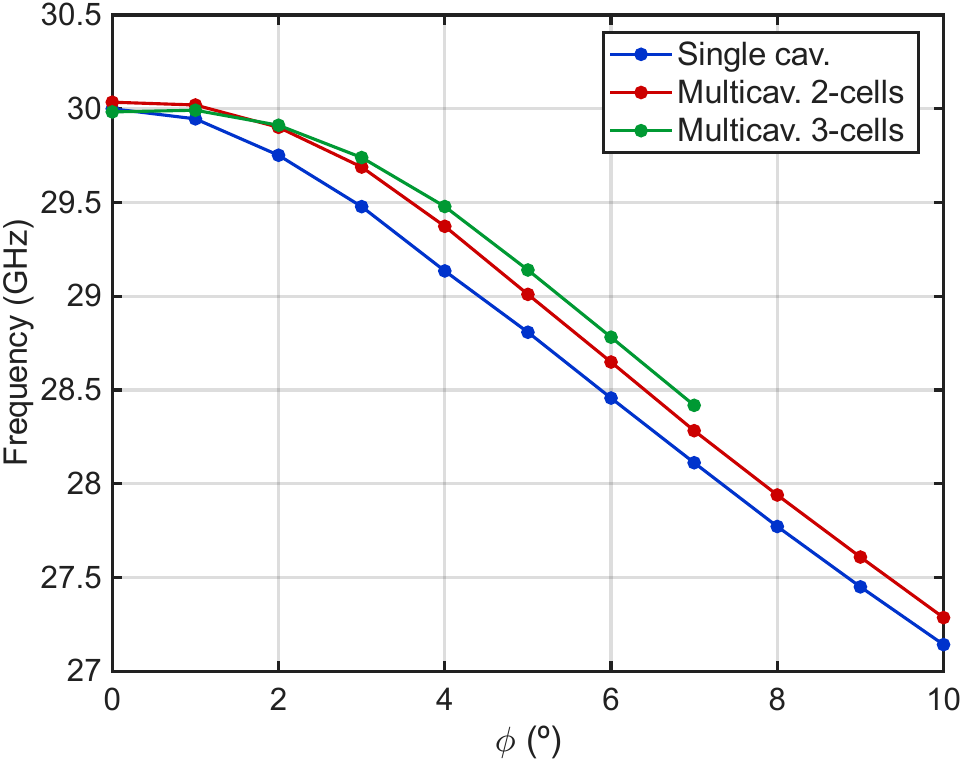}
         \caption{}
         \label{fig:Frequency}
\end{subfigure}
\hfill
\begin{subfigure}[b]{0.4\textwidth}
         \centering
         \includegraphics[width=1\textwidth]{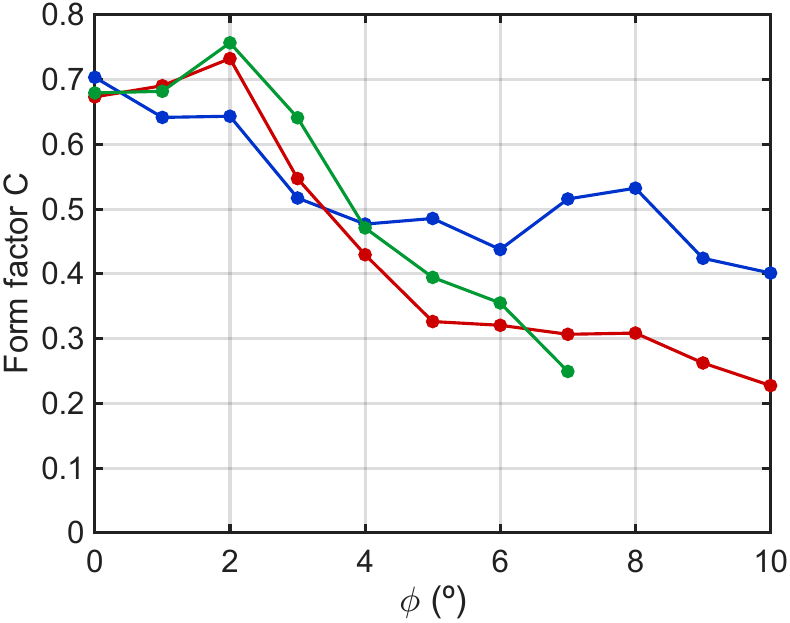}
         \caption{}
         \label{fig:FormFactor}
\end{subfigure}
\hfill
\begin{subfigure}[b]{0.4\textwidth}
         \centering
         \includegraphics[width=1\textwidth]{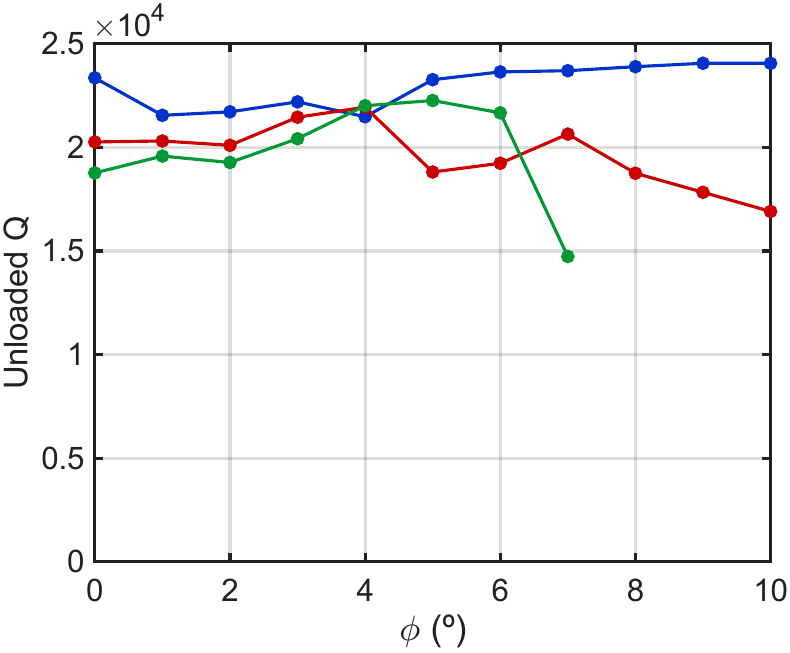}
         \caption{}
         \label{fig:UnloadedQ}
\end{subfigure}
\hfill
\begin{subfigure}[b]{0.4\textwidth}
         \centering
         \includegraphics[width=1\textwidth]{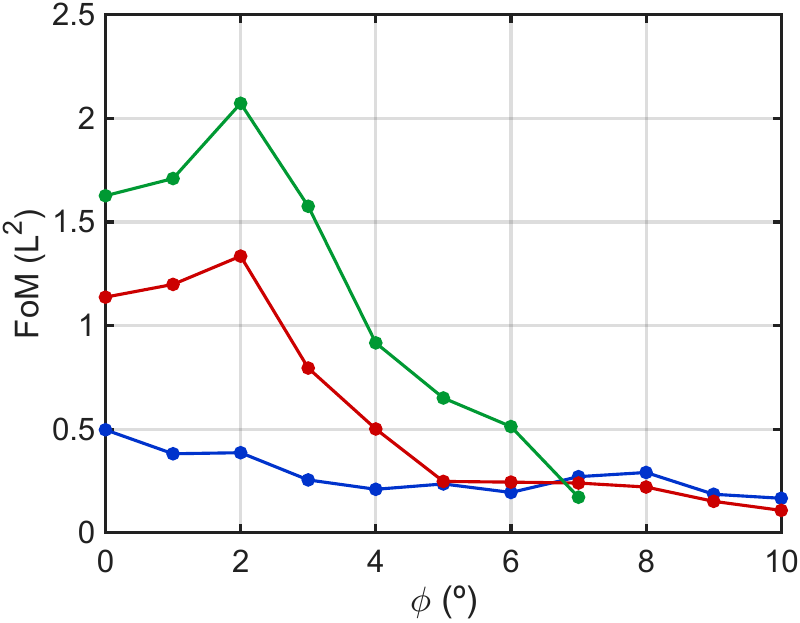}
         \caption{}
         \label{fig:FoM}
\end{subfigure}
\caption{Results from CST Frequency Domain simulations showing (a) resonant frequency of the $\mathrm{TM_{010}}$ mode, (b) form factor variation, (c) unloaded quality factor and (d) FoM versus rotation angle for the single coaxial cavity (blue line), the 2-cells multicavity (red line), and the 3-cells multicavity (green line).}
\label{fig:single_vs_multi}
\end{figure}

In \cite{QUAX_poly}, the studied tuning range ($0^\circ \leq \phi \leq 30^\circ$) shows a point beyond which the pseudo-$\mathrm{TM_{010}}$ mode approaches in frequency to other modes, giving rise to a mixed mode called $\mathrm{TM_{010}}$ family. For this reason, in this work, the rotation angle ($\phi$) of the internal prisms is limited by $0^\circ \leq \phi \leq 10^\circ$. After a set of CST Frequency Domain simulations with $1^\circ$ step, the frequency tuning range, the form factor, the unloaded quality factor and the FoM values are obtained for both designs, maintaining a port coupling close to $\beta = 2$, the desired value for optimising the scanning rate \cite{CAPP_revisiting}. The comparison of the outcomes versus the angle $\phi$ is shown in Figure~\ref{fig:single_vs_multi}.

\begin{figure}[htb]
\centering
\begin{subfigure}[b]{0.4\textwidth}
         \centering
         \includegraphics[width=1\textwidth]{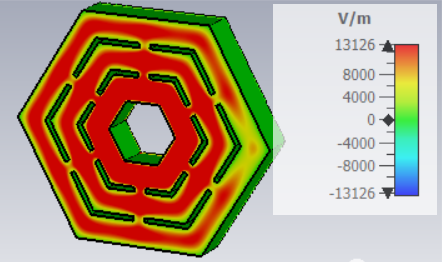}
         \caption{}
         \label{fig:Efields_Phi0}
\end{subfigure}
\hfill
\begin{subfigure}[b]{0.4\textwidth}
         \centering
         \includegraphics[width=1\textwidth]{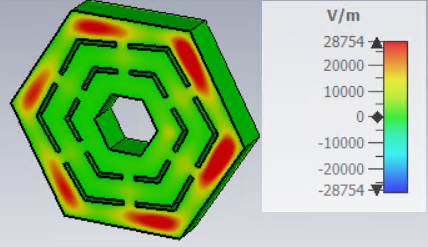}
         \caption{}
         \label{fig:Efields_Phi5}
\end{subfigure}
\caption{Longitudinal component of the axion mode electric field magnitude for (a) $\phi=0^\circ$, and (b) $\phi=5^\circ$. The 3D model is shown with a cross-sectional view at mid-height. The results are obtained in CST simulations.}
\label{fig:Efields_creating2DstructureMode}
\end{figure}

As it is depicted, the tuning range achieved by rotating the internal prism $\phi = 10^\circ$ is around $10$~$\%$ for all the designs\footnote{The relative tuning is calculated as $\frac{f_{max}-f_{min}}{f_c}$ being $f_c$ the central frequency of the tuning range.}. However, it can be seen that from $\phi = 5^\circ$ onwards, both the unloaded quality factor and the form factor decrease drastically in the multicavity design, affecting the FoM and restricting the tuning range of this design to $\phi_{max}=5^\circ$ and $\phi_{max}=7^\circ$ for the double and triple multicavity, respectively. This deterioration in performance may be due to a combination of factors related to mode accumulation at the vertices (caused by the geometry created during rotation) and the alteration of the interresonator coupling $k$, which causes the haloscope to shift from its optimal position. In fact, the rotation of the inner prisms converts the initial one-dimensional multicavity (radially coupled) into a two-dimensional multicavity, with couplings in radial and azimuthal directions, configuring a $2\times6$ and a $3\times6$ multicavities for 2-cells and 3-cells cases, respectively. Figure~\ref{fig:Efields_creating2DstructureMode} shows clearly this 2D multicavity structure created by changing from $\phi = 0^\circ$ to $5^\circ$.
It is important to note that the electric field magnitude in the inner and middle subcavities shown for this latter case (Figure~\ref{fig:Efields_Phi5}) seems to be zero, but it is high enough to yield form factor values such as those shown in Figure~\ref{fig:FormFactor}. However, it is shown this way in the figure solely because of the scale of the legend, in order to correctly visualise the behaviour of the mode localisation as $\phi$ changes.

In addition, at least two mode crossings between $4^\circ < \phi < 5^\circ$ and $6^\circ < \phi < 7^\circ$ have been observed in the double multicavity design, implying a loss of the axion mode at certain frequencies. Moreover, a decrease in the input coupling in both designs also limits the available tuning range, but this happens in regions where the $C$ and $Q_0$ factors have already been drastically compromised. This issue is also discussed in the next section.

\section{Tuning dynamics and port coupling study}
\label{s:Tuning_dynamics_and_port_coupling}

\subsection{Effects of synchronous vs. asynchronous tuning on sensitivity}\label{ss:Synchronous_vs_asynchronous}

To evaluate the impact of the tuning dynamics on the triple multicavity sensitivity, both synchronous ($\phi_1 = \phi_2$, being $\phi_2$ the angle of the most inner prism) and asynchronous ($\phi_1 \neq \phi_2$) rotation regimes of the internal prisms were investigated. EM simulations revealed that asynchronous tuning drastically degrades the cavity form factor ($C$). Even a minimal phase difference of $1^\circ$ between the prisms disrupts the symmetry of the resonant mode's electric field, leading to a severe reduction in its overlap with the external static magnetic field. Consequently, this configuration compromises the axion-photon conversion power. Due to this poor performance in the form factor from the very first degrees of desynchronisation, the asynchronous tuning approach was discarded, establishing strict synchronous rotation as a primary operational requirement.

Following this, to ensure an efficient frequency sweep, the design of the two multicavities was structurally optimised for an intermediate synchronous tuning angle of $\phi = 2.5^\circ$. This intermediate-angle optimisation aims to achieve a more balanced and effective frequency tuning profile, maximising the tuning range while preventing the severe degradation of both the form factor and the unloaded quality factor ($Q_0$) across the operational band. However, it has been observed that the results are similar to the original design and tuning operation.

\subsection{Port coupling optimisation for signal extraction}\label{ss:Port_coupling}

\begin{figure}[htb]
\centering
\begin{subfigure}[b]{0.4\textwidth}
         \centering
         \includegraphics[width=1\textwidth]{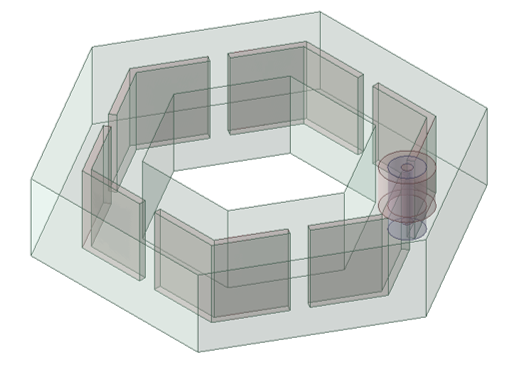}
         \caption{}
         \label{fig:Dual_PortAtExternalCav}
\end{subfigure}
\hfill
\begin{subfigure}[b]{0.4\textwidth}
         \centering
         \includegraphics[width=1\textwidth]{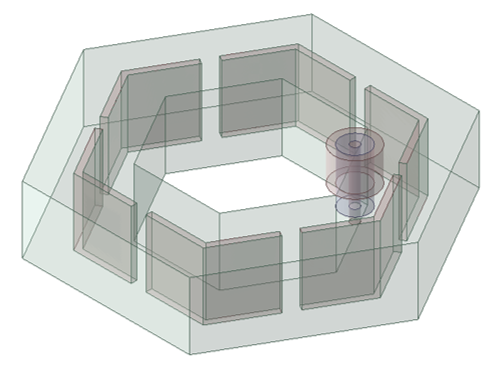}
         \caption{}
         \label{fig:Dual_PortAtInternalCav}
\end{subfigure}
\hfill
\begin{subfigure}[b]{0.32\textwidth}
         \centering
         \includegraphics[width=1\textwidth]{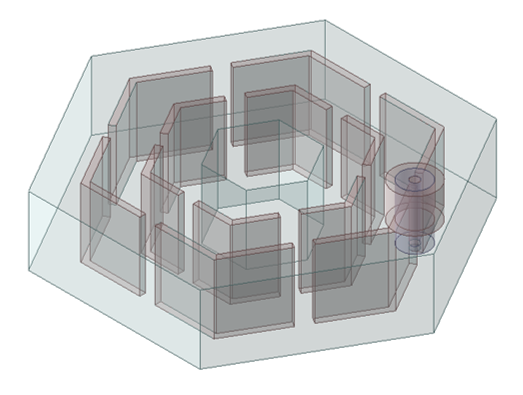}
         \caption{}
         \label{fig:Triple_PortAtExternalCav}
\end{subfigure}
\hfill
\begin{subfigure}[b]{0.32\textwidth}
         \centering
         \includegraphics[width=1\textwidth]{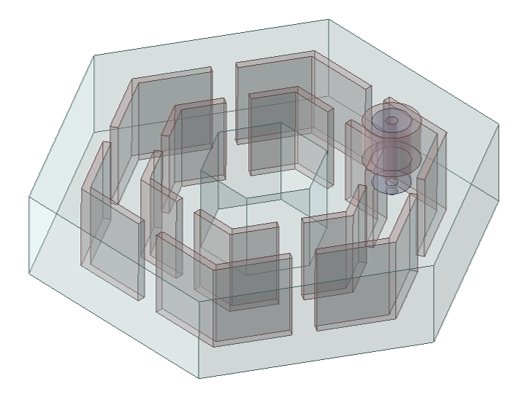}
         \caption{}
         \label{fig:Triple_PortAtCentralCav}
\end{subfigure}
\hfill
\begin{subfigure}[b]{0.32\textwidth}
         \centering
         \includegraphics[width=1\textwidth]{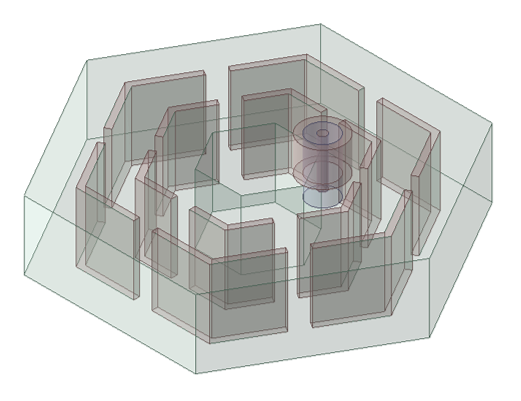}
         \caption{}
         \label{fig:Triple_PortAtInternalCav}
\end{subfigure}
\caption{3D model of the tested configurations, showing the port positioned in the subcavities: (a) external and (b) internal for the double multicavity; and (c) external, (d) central, and (e) internal for the triple multicavity.}
\label{fig:PortAtDifferentSubcavs}
\end{figure}

Efficient extraction of the signal generated by axion-photon conversion requires rigorous control over the coupling between the cavity’s resonant mode and the amplification chain. Consequently, the port coupling mechanism represents a critical technical constraint that dictates the overall design of the haloscope. Observations throughout the characterisation of these multicavity structures indicate that maintaining a perfectly symmetric and optimised RF electric field pattern across all subcavities is unfeasible during the tuning process and, indeed, even in the initial configuration. Due to this dynamic field asymmetry, the local maxima of the RF electric field shift and concentrate within different regions of the structure. To achieve the target port coupling factor, it is imperative to exploit this behaviour by positioning the extraction port in the subcavities where the field intensity is most favourable. Figure~\ref{fig:PortAtDifferentSubcavs} shows the different configurations that have been tested to achieve a $\beta$ value of $2$ at the port.

Specifically, to maintain the desired overcoupled regime of $\beta = 2$, the response of both the 2-cells and 3-cells designs was evaluated as a function of the tuning angle ($\phi$). It was found that for angles $\phi \geq 5^\circ$, the extraction port must be repositioned to a more external subcavity in both designs. The underlying physical reason for this requirement is the spatial evolution of the mode: as the internal prisms rotate and $\phi$ increases, the electric field pattern in the innermost cavities undergoes rapid distortion, losing its alignment with the coupling antenna significantly faster than in the outer regions. Relocating the port to an outer subcavity compensates for this local degradation, ensuring that the antenna maintains sufficient interaction with the electric field to sustain $\beta = 2$ at higher tuning angles.

\section{Scaling potential}
\label{s:Scaling_potential}

\subsection{Extension to Quad-cells multicavity}\label{ss:Quad}

To evaluate the scalability potential of the proposed system, the theoretical and technical feasibility of implementing a quad-subcavity architecture has been investigated. The primary challenge in this geometric expansion lies in the strict spatial constraints imposed by the experimental environment, specifically, the magnet bore diameter, which limits the maximum outer radius of the structure to $r_{out} = 25$~mm. Furthermore, any topological modification must ensure that the operational resonant frequency remains consistent with the values established in this work.

During the conceptual design phase of this quadruple configuration, significant difficulties emerged due to these volume restrictions. Two primary geometric approaches were evaluated to accommodate the fourth subcavity: the inclusion of an internal coaxial cavity and the development of a hexagonal cavity (omitting the internal prism). In both scenarios, the parametric margin for optimising the RF fields is severely restricted.

In particular, the analysis of the hexagonal cavity topology reveals a critical trade-off in radial packing. Design calculations indicate that a minimum space of $10.6$~mm is required (comprising $8.6$~mm for the core structure plus 2 mm for separation or coupling requirements). Given that the maximum available radial space in the cross-section is $12.32$~mm, the implementation is dimensionally feasible. To address the resulting central clearance and achieve the target frequency tuning without compromising structural integrity, a strategic increase in the cavity wall thickness is proposed as an engineering solution. This adjustment would compensate for the volumetric gap, ensuring adequate EM field confinement within the strict $25$~mm boundary.

\begin{figure}[htb]
\centering
\begin{subfigure}[b]{0.47\textwidth}
         \centering
         \includegraphics[width=1\textwidth]{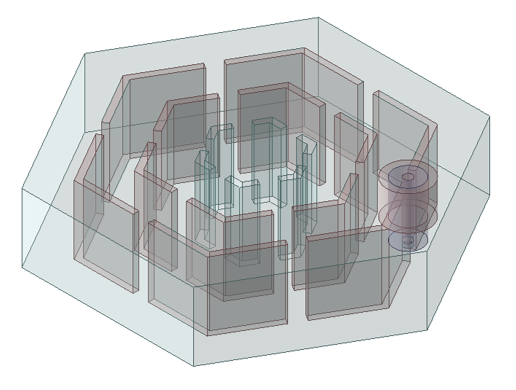}
         \caption{}
         \label{fig:QuadCav_EmptyCav}
\end{subfigure}
\hfill
\begin{subfigure}[b]{0.49\textwidth}
         \centering
         \includegraphics[width=1\textwidth]{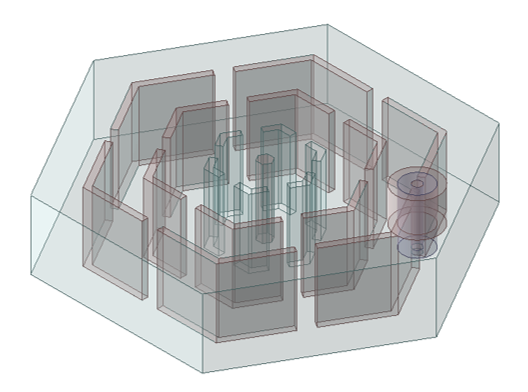}
         \caption{}
         \label{fig:QuadCav_CoaxCav}
\end{subfigure}
\caption{3D model of a 4-cell haloscope structure with a differently shaped innermost cavity: (a) with an empty hexagonal cavity and (b) with a coaxial hexagonal cavity.}
\label{fig:QuadCav}
\end{figure}

This implementation has been proposed as a future development and would be based on one of the two geometries shown in Figures~\ref{fig:QuadCav_EmptyCav} and \ref{fig:QuadCav_CoaxCav}.

\subsection{Practical challenges in high-order multicavity systems}\label{ss:Practical_challenges}

The transition toward higher-order systems represents a promising path for leveraging most of the magnet bore space and thus enhancing the performance of axion detectors. To fully exploit the advantages of a 4-cells structure, certain technical refinements must be integrated into the design to ensure peak operational efficiency:

\begin{itemize}

    \item High-precision manufacturing: As the design approaches the structural limits of the $r_{out} = 25$~mm bore, the available central clearance offers an opportunity for advanced high-precision machining. Maintaining strict tolerances is essential not as a barrier, but as a means to ensure that the increased geometric complexity translates into the predicted theoretical sensitivity, especially under the demanding conditions of cryogenic environments.
    \item Advanced spectral control and tuning: The increased complexity of high-order systems needs a more sophisticated control of resonant modes. While the alignment of four simultaneous resonances requires a more refined tuning strategy than in triple-cavity systems, overcoming this challenge avoids the mode clustering.
    \item Enhanced thermal stabilisation: The higher surface-to-volume ratio inherent in multicavity designs provides a larger area for potential thermal interaction. By implementing optimised thermal anchoring strategies, this characteristic can be leveraged to maintain high-Q factors and exceptional frequency stability, ensuring reliable data acquisition during extended axion experimental runs.
\end{itemize}

\section{Conclusions and future outlook}
\label{s:Conclusions}

In this work, a comprehensive study on the implementation of a multicavity approach applied to hexagonal coaxial geometries for axion search experiments has been presented. By transitioning from a baseline single-cavity design to double and triple-subcavity configurations, it has been demonstrated that it is possible to significantly enhance the detector's performance while operating within strict volume constraints and keeping only one extraction point, avoiding the coherent sum needed in a multiple cavity concept.

The performance analysis indicates that the multicavity architectures achieve a maximum Figure of Merit of $2.073$~$\mathrm{L}^2$, with an average improvement of $\times\sim3$ over the tuning range in the triple multicavity case compared to the single cavity baseline. Furthermore, the inner prism rotation system has proved to be an effective tuning mechanism, providing a continuous spectral coverage of $1.565$~GHz ($5.2$~$\%$) with a frequency sensitivity of $-358.4$~MHz$/^\circ$. The study of tuning dynamics confirms that maintaining synchronicity between subcavities is feasible, although it requires a high degree of mechanical precision to avoid sensitivity degradation due to mode splitting. As for the practical limitations of these designs, these are primarily due to two factors. On one hand, there are certain rotation angles of the internal prism at which operation is not possible due to mode crossings, which significantly reduce the form factor and the quality factor. On the other hand, it has been found that beyond a certain rotation angle, port coupling decreases, which hinders the optimisation of the scanning rate. Despite these drawbacks, for a limited tuning range, multicavity designs offer superior performance to single-cavity designs.

Regarding the scaling potential, it has been shown that a quad-subcavity configuration is theoretically viable even under the tight constraints of a $25$~mm bore radius. The proposed engineering solution, adjusting wall thicknesses to optimise a $10.6$~mm structure within a $12.32$~mm radial space, allows for a robust design that preserves the target resonant frequency. While higher-order systems introduce increased complexity in terms of manufacturing tolerances and thermal anchoring, the potential gains in sensitivity justify the pursuit of these advanced architectures. In this way, higher frequencies could be explored with the same limited space by increasing the number of cells.

At this point, short cavities ($10$~mm long) have been designed due to computational constraints. In order to leverage the length of the magnet bore, longer structures will be simulated in the future, establishing the bounds related with the form factor degradation and coupling feasibility. 

Future work will also focus, firstly, on designing and simulating a longer multicavity haloscope to take advantage use of the magnet bore length, which presents a challenge in terms of computational cost. The next step is to proceed to the experimental validation of these designs. The fabrication of a precision-machined prototype for the double and triple-subcavity systems is the next logical step to characterise the actual Q-factor and tuning reliability at cryogenic temperatures. These advancements pave the way for a new generation of high-sensitivity axion haloscopes capable of proving not previously explored regions of DM axion masses.

\acknowledgments
This work was performed within the RADES group; we thank our colleagues for their support. The work of JMGB was supported by the European Research Council grant ERC-2018-StG-802836 (AxScale project), the Lise Meitner program 'In search of a new, light physics' of the Max Planck Society, as well as by Deutsche Forschungsgemeinschaft (DFG) through Grant No. 532766533. This research was also supported under the Horizon Europe programme (ERC-2023-SyG DarkQuantum, grant agreement No. 101118911). The publication is part of the Quantera QRADES project: action PCI2024-153487, funded by MCIU/AEI/10.13039/501100011033 and co-funded by the European Union. The research leading to these results has received funding from the Spanish Ministry of Science and Innovation with the project PID2022-137268NBC53 funded by MICIU/AEI/10.13039/501100011033/ and by “ERDF/EU”. This article is based upon work from COST Action COSMIC WISPers CA21106, supported by COST (European Cooperation in Science and Technology).

\bibliographystyle{JHEP.bst}
\bibliography{mybibfile.bib}
\end{document}